\documentclass[reprint,superscriptaddress,nofootinbib,twocolumn,amsmath,amssymb,aps,prl,longbibliography]{revtex4-2}

\usepackage{graphicx}% Include figure files
\usepackage{bm}% bold math
\usepackage[colorlinks]{hyperref}

\newcommand{\bra}[1]{\left< #1 \right\vert}
\newcommand{\ket}[1]{\left\vert #1 \right>}

\newcommand{\braket}[1]{\langle#1\rangle}
\DeclareMathOperator{\sinc}{sinc}

\begin{document}

\preprint{APS/123-QED}

\title{Observation of anomalous classical-to-quantum transitions in many-body systems}

\author{C. You}
\affiliation{Quantum Photonics Laboratory, Department of Physics \& Astronomy, Louisiana State University, Baton Rouge, LA 70803, USA}

\author{M. Hong}
\affiliation{Quantum Photonics Laboratory, Department of Physics \& Astronomy, Louisiana State University, Baton Rouge, LA 70803, USA}

\author{F. Mostafavi}
\affiliation{Quantum Photonics Laboratory, Department of Physics \& Astronomy, Louisiana State University, Baton Rouge, LA 70803, USA}

\author{J. Ferdous}
\affiliation{Quantum Photonics Laboratory, Department of Physics \& Astronomy, Louisiana State University, Baton Rouge, LA 70803, USA}

\author{R. de J. Le\'on-Montiel}
\affiliation{Instituto de Ciencias Nucleares, Universidad Nacional Aut\'onoma de M\'exico, Apartado Postal 70-543, 04510 Cd. Mx., M\'exico}

\author{R. B. Dawkins}
\email{rdawki1@lsu.edu}
\affiliation{Quantum Photonics Laboratory, Department of Physics \& Astronomy, Louisiana State University, Baton Rouge, LA 70803, USA}

\author{O. S. Maga\~na-Loaiza}
\affiliation{Quantum Photonics Laboratory, Department of Physics \& Astronomy, Louisiana State University, Baton Rouge, LA 70803, USA}

\date{\today}

\begin{abstract}

The correspondence principle bridges the quantum and classical worlds by establishing a direct link between their dynamics. This well-accepted tenant of quantum physics has been explored in quantum systems wherein the number of particles is increased to macroscopic scales. However, theoretical investigations of nanoscale structures have revealed discrepancies when attempting to bridge classical and quantum physics. Here, we report on the experimental observation of anomalous classical-to-quantum transitions in open many-body optical systems. We demonstrate, for the first time, the lack of classical-to-quantum correspondence between a macroscopic optical system and its constituent quantum multiphoton subsystems.  In contrast to common belief, we demonstrate that the coherence dynamics of many-body quantum subsystems with up to forty particles can indeed be opposite to that exhibited by the hosting macroscopic system.  By employing complex-Gaussian statistics, we show that these effects are universal for open many-body systems. Consequently, our work can have important implications for other fields of physics ranging from condensed matter to nuclear physics. 

\end{abstract}

\maketitle

The correspondence principle stipulates that the quantum mechanical description of macroscopic systems, comprising large number of particles $n$, approaches the predictions of classical physics \cite{ZurekPT, MartiniRMP, JaegerAJP}. This fundamental principle has been explored by approximating the conditions of quantum physical systems to those that characterize their classical counterparts \cite{JarzynskiPRX2015, EstebanNatCom2012}. In this regard, the quantum theory of electromagnetic radiation indicates that light emitted by a blackbody converges to the classical Rayleigh-Jeans law for long wavelengths \cite{MehtaPR1964, WangPRL2022}. In this case, the equivalence between the quantum and classical theories of the light field implies that the Planck constant $\hbar$ vanishes for blackbody radiation characterized by long wavelengths \cite{MehtaPR1964, WangPRL2022, KasimirProgOpt2017}. These situations, in which $n \rightarrow \infty$ and $\hbar \rightarrow 0$, represent the limits of the Planck-Bohr correspondence principle \cite{JaegerAJP, LiboffPhysToday1984}. Recently, the transition between the so-called classical and quantum worlds have been theoretically investigated in subwavelength photonic structures and photonic circuits \cite{EstebanNatCom2012, SavageNature2012, HongNP2024, JarzynskiPRX2015, MalyNature2023}. However, these studies unveiled discrepancies when attempting to bridge classical and quantum physics. Nevertheless, the correspondence between classical macroscopic systems and their constituent many-body quantum components remains unexplored.

\begin{figure*}[!t]
  \centering
 \includegraphics[width=0.99\textwidth]{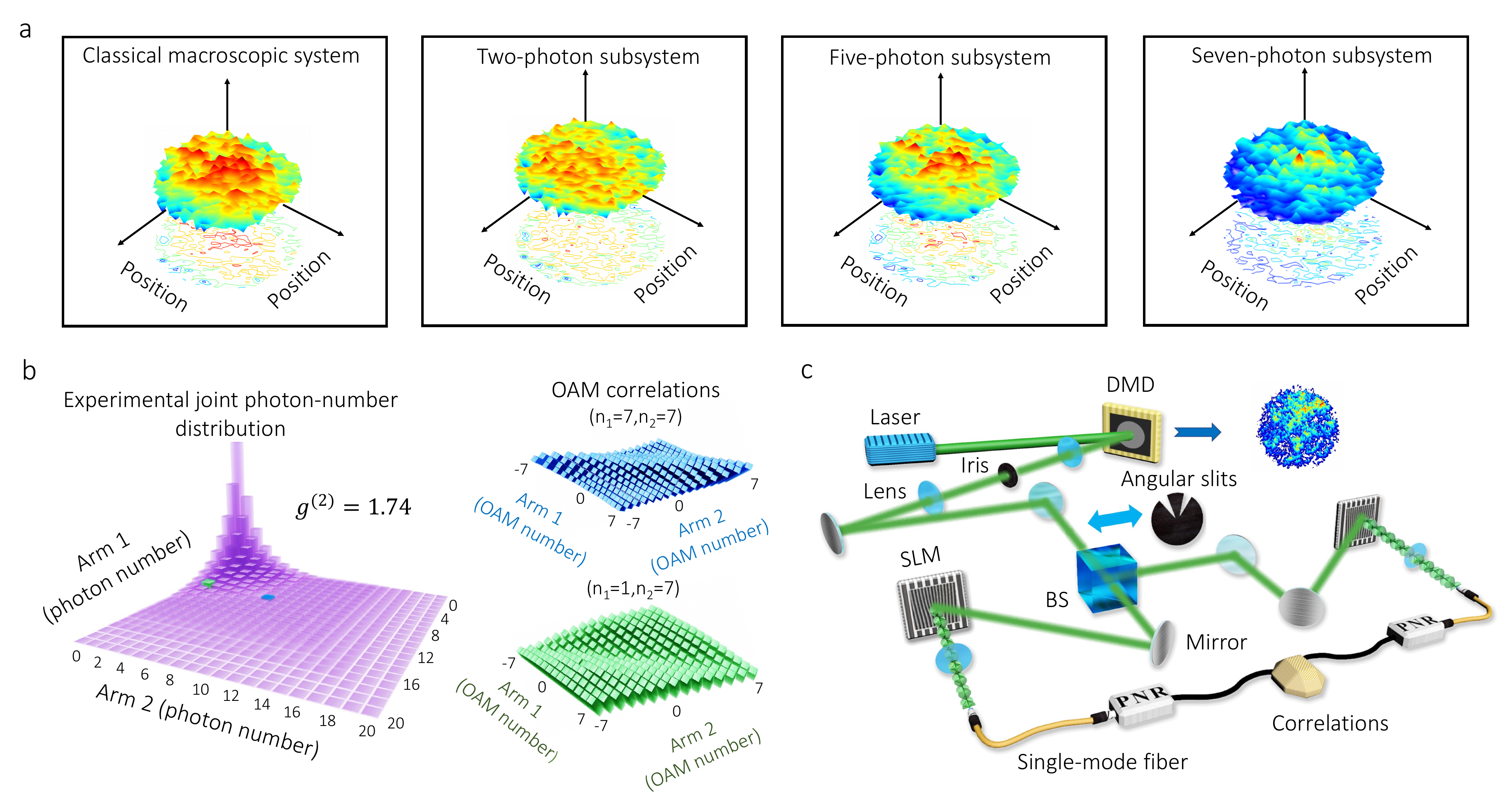}
\caption{\textbf{Open macroscopic optical system and its constituent quantum many-body subsystems.} The classical macroscopic system in our experiment consists of a random optical field with transverse spatial singularities \cite{BerryJPA1978, BenderOptica2021}. This is reported in the first panel of \textbf{a}. Interestingly, the optical macroscopic system hosts many-body quantum subsystems with different distributions of vortices.  Our experimental measurement of the spatial distribution of the many-body subsystems is reported in the other panels of \textbf{a}. We extract the quantum multiparticle subsystems from the open macroscopic system by implementing projective measurements on Fock states \cite{Fatemeh2024Imaging}. In panel \textbf{b}, we report the experimentally measured joint photon-number distribution of our macroscopic system. This shows our capability to isolate many-body quantum subsystems with up to forty particles. The degree of second-order coherence $g^{(2)}$ for the macroscopic system is 1.74, which corresponds to the degree of coherence of a classical light beam \cite{MandelBook}. The OAM correlations for two projected many-body subsystems consisting of eight and fourteen particles are shown in the right part of \textbf{b}. These correlations are measured using the setup depicted in \textbf{c}. Here we use a digital micromirror device (DMD) to produce the macroscopic optical systems containing multiple independent wavepackets with randomly modulated phases \cite{SmithPRL,HongLPR}.  We pass the beam through a beam splitter to measure its correlation properties. We add an angular double slit to study its quantum multiparticle dynamics. The beams reflected by the spatial light modulators (SLMs) are projected onto specific OAM values and then measured by two superconducting nanowire single-photon detectors that are used to perform photon-number-resolving (PNR) detection \cite{omar2019Multiphoton, YouAPR2020, YouAPR2021}. Our experiment is finalized by correlating the signals from the detectors. This kind of measurement enables the isolation of specific quantum dynamics.}
\label{Fig1}
\end{figure*}

Here we isolate the quantum many-body components within an open classical macroscopic optical system \cite{HarderPRL2016, SmithPRL, HongLPR, GattiPRL2003, AnnoPhysRep}. This capability enables us to demonstrate the lack of a direct correspondence between a classical physical system and its quantum constituents. This surprising observation deviates from common belief and interpretations of the correspondence principle in quantum physics \cite{ZurekPT, MartiniRMP, JaegerAJP}. In our work, we extract multiphoton quantum subsystems by performing projective measurements of the particle number of a classical optical system \cite{omar2019Multiphoton, YouAPR2020, YouAPR2021}. This measurement scheme enables us to observe the quantum coherence of many-body subsystems with up to forty interacting particles. Specifically, we explore classical-to-quantum transitions by measuring multiparticle coherence of vortices produced by the interference of random classical light waves \cite{BerryJPA1978, DennisProgOpt2009, gbur2016singular, MaganaSciAdv}.  Interestingly, we demonstrate that the extracted many-body systems exhibiting anti-coalescence effects induce correlated interactions. Furthermore, we found that those multiparticle subsystems exhibiting coalescence produce anti-correlated dynamics, which are opposite to the macroscopic behavior of the classical hosting system \cite{JeltesNature2007}. Remarkably, the potential to isolate quantum coherence within classical open systems not only holds fundamental significance for quantum physics but also presents profound implications for retrieving the quantum features of open physical systems affected by decoherence processes \cite{WubsNatPhy2024, Fatemeh2024Imaging, MatthijsScience2008, TingScience2009}.

The formation of dislocations in incoherent random waves has sparked significant interest in the investigation of vortices in diverse physical systems \cite{BerryRSLA1974, BerryJPA1978, DennisProgOpt2009, BenderOptica2021, gbur2016singular, WrightPRL2009}.  In this work, we prepared a classical random optical system with many transverse spatial singularities. This is shown in the first panel of Figure \ref{Fig1}\textbf{a} \cite{HongLPR}. This classical system hosts many-body quantum subsystems whose vortex dynamics are nonclassical and cannot be described using Berry's formalism \cite{BerryRSLA1974, BerryJPA1978, gbur2016singular}. Remarkably, the constituent many-body subsystems within the classical random system unveil different spatial structures with distinct transverse vortex distributions. We extract multiphoton quantum subsystems by performing projective measurements on Fock states \cite{GaluberPR}. As described below, these measurements enable us to collapse the state of the scattered macroscopic system onto isolated multiphoton subsystems \cite {Fatemeh2024Imaging}. Interestingly, the extracted multiparticle subsystems reported in Figure \ref{Fig1}\textbf{a} exhibit peculiar vortex dynamics. The presence of these vortices imprints an orbital angular momentum (OAM) spectrum onto our macroscopic system \cite{gbur2016singular, MaganaSciAdv}. Here, we characterize its vortex dynamics by analyzing the OAM correlations of its constituent multiparticle subsystems. This can be achieved by expanding the following OAM-dependent density matrix in the Fock state basis, which we have here written in the coherent-state basis for simplicity as 
\begin{equation} 
\label{Eq1}
    \hat{\rho}_{\ell_1,\ell_2} = \int d^2\alpha d^2\beta  P_{\ell_1,\ell_2}(\alpha,\beta)|\alpha,\beta\rangle\langle\alpha,\beta|,
\end{equation}
where 
\begin{widetext}
    \begin{equation}
    \label{Eq2}
    P_{\ell_1,\ell_2}(\alpha,\beta) = \frac{1}{4\pi^2 \big|\sigma_{\ell_1}\sigma_{\ell_2} - |\eta_{\ell_1,\ell_2}|^2\big|} \exp\left[-\frac{\sigma_{\ell_2}|\alpha - \mu_{\ell_1}|^2 + \sigma_{\ell_1}|\beta - \mu_{\ell_2}|^2 - 2\text{Re}\left[(\alpha^*-\mu_{\ell_1}^*)(\beta-\mu_{\ell_2})\eta_{\ell_1,\ell_2}\right]}{2(\sigma_{\ell_1}\sigma_{\ell_2} - |\eta_{\ell_1,\ell_2}|^2)}\right].
\end{equation}
\end{widetext}

\begin{figure*}[!t]
  \centering
 \includegraphics[width=0.99\textwidth]{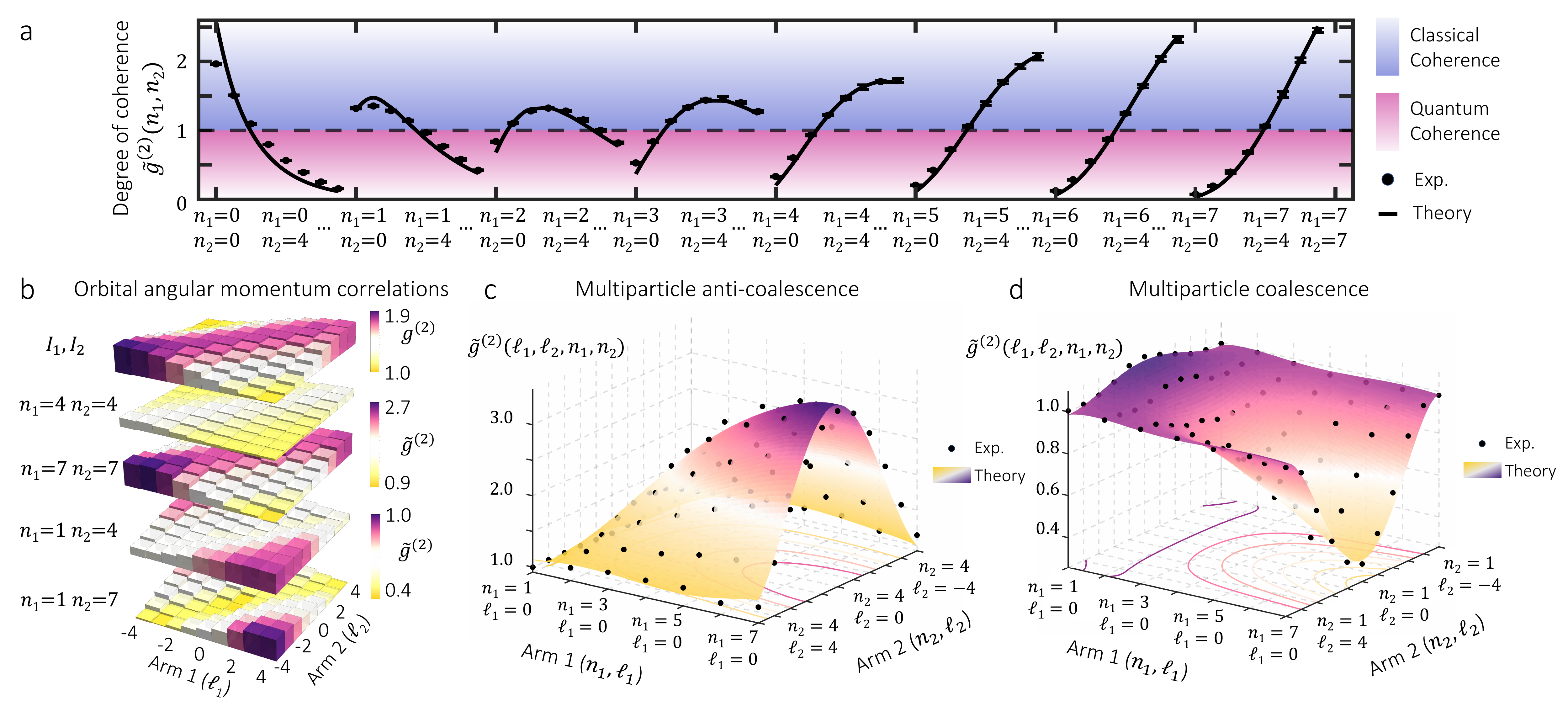}
\caption{\textbf{Classical-to-quantum coherence transitions of many-body subsystem.}  As indicated in \textbf{a}, the extracted multiparticle subsystems can exhibit either classical ($\tilde{g}^{(2)}(n_1,n_2)>1$) or quantum ($\tilde{g}^{(2)}(n_1,n_2)<1$) properties of coherence.
Interestingly, these properties vary with the number of particles in the extracted many-body systems as $n \rightarrow \infty$. In this case, we are reporting multiparticle systems that do not carry OAM. Despite the many multiphoton subsystems exhibiting quantum coherence, most of the multiphoton subsystems exhibit classical properties of coherence. Indeed, the quantum dynamics hosted by the macroscopic system are overwhelmed by the classical properties of coherence of the majority of the constituent subsystems. The theoretical predictions were calculated based on Eqs. (\ref{Eq3}) and (\ref{Eq4}). Furthermore, we report the classical OAM correlations of the random macroscopic system in the first row of \textbf{b}. These correlations are quantified through the second-order coherence function $g^{(2)}(\ell_1,\ell_2)$. The second and third rows therein report similar correlation peaks for symmetric multiparticle subsystems with eight ($n_1=4,n_2=4$) and fourteen ($n_1=7,n_2=7$) photons. These correlations are quantified through the coherence function $\tilde{g}^{(2)}(\ell_1,\ell_2,n_1,n_2)$. Remarkably, the fourth and fifth rows in \textbf{b} demonstrate the existence of OAM-correlation holes for asymmetric subsystems with five ($n_1=1,n_2=4$) and eight  ($n_1=1,n_2=7$) photons. Interestingly, these processes lead to generalized forms of multiparticle interference that we explore using the angular double-slit structure in Fig. \ref{Fig1}\textbf{c}. As reported in \textbf{c}, we observe multiparticle Hanbury Brown and Twiss interference as we scan the OAM of the extracted subsystems with $\tilde{g}^{(2)}(\ell_1,\ell_2,n_1,n_2)>1$ \cite{MaganaSciAdv}. Notably, the many-body subsystems characterized by $\tilde{g}^{(2)}(\ell_1,\ell_2,n_1,n_2)<1$ exhibit a type of multiparticle Hong-Ou-Mandel effect as shown in \textbf{d} \cite{HiekkamakiPRL2021}.}
\label{Fig2}
\end{figure*}

The function $P_{\ell_1,\ell_2}(\alpha,\beta)$ is a complex-Gaussian probability density function for our classical random optical system \cite{dawkins2024quantum}. Here, $\alpha$ and $\beta$ are the coherent amplitudes of the two OAM modes $\ell_1$ and $\ell_2$. Furthermore, the average electric-field amplitudes are represented by $\mu_{\ell_i}$, and $\sigma_{\ell_i}$ and $\eta_{\ell_i,\ell_j}$ are the respective variances and covariances of these electric-field amplitudes. More detailed information can be found in the supplementary information (SI).

Our classical macroscopic system, described by Eqs. (\ref{Eq1}) and  (\ref{Eq2}), hosts many multiparticle quantum systems. In Figure \ref{Fig1}\textbf{b}, we report the experimental joint photon-number distribution of our system hosting many-body subsystems with up to forty photons. Given the classical nature of our macroscopic random optical field, this can be fully described using Maxwell's equations \cite{MandelBook,MaganaSciAdv,BenderOptica2021}. The differences among the spatial structures of the multiparticle subsystems in Figure \ref{Fig1}\textbf{a} produce distinct OAM correlations. As shown in Figure \ref{Fig1}\textbf{b}, the strength and nature of these correlations are defined by the isolated many-body quantum subsystem \cite{MalyNature2023}. The origin of these correlations is described below, and these measurements were performed using the experimental setup shown in Figure \ref{Fig1}\textbf{c}. This setup allowed us to measure OAM correlations for the multiparticle subsystems within the macroscopic random optical system. The use of photon-number-resolving (PNR) detection enabled us to collapse the state of the scattered macroscopic system into isolated multiparticle subsystems that exhibit different quantum dynamics \cite{Fatemeh2024Imaging, YouAPR2021, BhusalNpj}. The projection of the classical macroscopic system into isolated subsystems unveils peculiar quantum dynamics which can be opposite to the classical dynamics of the hosting system \cite{ZurekPT,MartiniRMP,ArndtNatPhys,JaegerAJP}.

\begin{figure*}[!ht]
  \centering
 \includegraphics[width=0.99\textwidth]{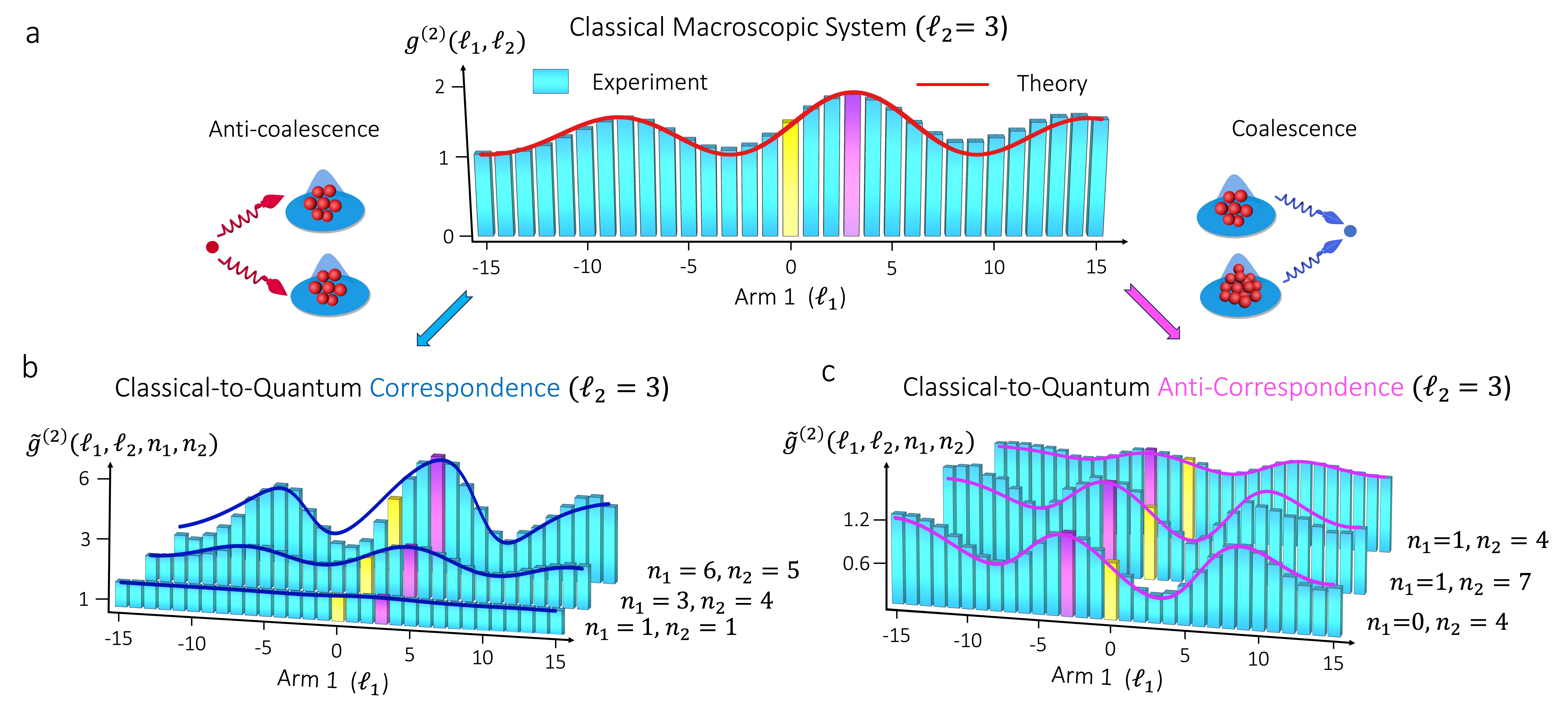}
\caption{\textbf{Observation of ordinary and anomalous classical-to-quantum transitions in many-body systems.} The scattering of our random macroscopic optical system by a double-slit structure produces the classical interference pattern shown in \textbf{a}. This pattern was obtained by projecting the scattered random field in one of the arms into different OAM modes, from $\ell_1=-15$ to $\ell_1=15$, whereas the scattered photons in the second arm are projected into $\ell_2=3$. The separation between the slits is $\pi/6$ radians and their widths are $\pi/12$ radians. The overwhelming presence of multiparticle subsystems with $\tilde{g}^{(2)}(n_1,n_2)>1$ in Figure \ref{Fig2} induce a correlation shift of three units to the right.  These kinds of many-body subsystems exhibiting anti-coalescence dynamics produce other correlated structures with a peak at $\ell_1=3$. We report these measurements in \textbf{b}. Interestingly, the visibility of these interference patterns increases with the number of particles in the subsystem, enabling a quantum-enhanced visibility for the many-body subsystem with eleven particles $(n_1 = 6, n_2 = 5 )$. Despite the direct correspondence between the interference structures in \textbf{a} and \textbf{b}, there are quantum processes of coherence inducing the increasing visibilities in \textbf{b} that surpass that of the classical hosting system in \textbf{a}. Remarkably, the projection of the classical macroscopic system into the many-body subsystems in \textbf{c} unveils a classical-to-quantum anti-correspondence. The coalescence dynamics hosted by these kinds of subsystems, characterized by $\tilde{g}^{(2)}(n_1,n_2)<1$, produce anti-correlated interference patterns with a peak at $\ell_1=-3$.  Even though the isolated quantum subsystems in \textbf{c} were extracted from the open classical system in \textbf{a}, they lack a direct classical-to-quantum correspondence. }
\label{Fig3}
\end{figure*}

We now discuss the coherence properties of the macroscopic system and its constituent many-body subsystems. Given the classical nature of our macroscopic system, some of its properties can be characterized through the measurement of intensity correlations \cite{GaluberPR,MandelBook}. Specifically, the OAM correlation measurements can be described by projecting our macroscopic system onto the OAM modes $\ell_{1}$ and $\ell_{2}$. The density matrix in Eq. (\ref{Eq1}) allows us to express the degree of second-order coherence for our random hosting system as 
\begin{equation}
\label{Eq3}
    g^{(2)}(\ell_1,\ell_2) = 1 + \frac{4|\eta_{\ell_1\ell_2}|^2 + 4 \text{Re}\left[\mu^*_{\ell_1}\mu_{\ell_2}\eta_{\ell_1\ell_2}\right]}{(2\sigma_{\ell_1}+|\mu_{\ell_1}|^2)(2\sigma_{\ell_2}+|\mu_{\ell_2}|^2)}.
\end{equation} 
This expression certifies the classicality of the macroscopic system reported in Figure \ref{Fig1}\textbf{b}. The detailed derivation of Eq. (\ref{Eq3}) can be found in the SI. Despite the classical nature of our photonic system, we show that this macroscopic behavior results from the collective dynamics of nonclassical many-body subsystems. Notably, one can access their quantum dynamics by projecting the random optical field into its constituent multiparticle subsystems \cite{omar2019Multiphoton, YouAPR2020, YouAPR2021}. The coherence of these isolated subsystems can be described through the following coherence function
\begin{widetext}
\begin{equation}
\label{Eq4}
    \tilde{g}^{(2)}(\ell_1,\ell_2,n_1,n_2) = \frac{\text{Tr}\big[\hat{\rho}_{\ell_1,\ell_2}\ket{n_1,n_2}\bra{n_1,n_2}\big]}{\left(\sum_n \text{Tr}\big[\hat{\rho}_{\ell_1,\ell_2}\ket{n,n_2}\bra{n,n_2}\big]\right)\left(\sum_m \text{Tr}\big[\hat{\rho}_{\ell_1,\ell_2}\ket{n_1,m}\bra{n_1,m}\big]\right)}.
\end{equation}
\end{widetext}

Here, the coherence function $\tilde{g}^{(2)}(\ell_1,\ell_2,n_1,n_2)$ allows identifying the classical or quantum processes of coherence of the multiparticle subsystems that form the hosting random optical field \cite{HongNP2024, JeltesNature2007}. As shown in Figure \ref{Fig2}\textbf{a}, the properties of coherence $\tilde{g}^{(2)}(n_1,n_2)$ of the extracted subsystems oscillate as the number of particles increase ($n\rightarrow\infty$). Despite the many multiparticle subsystems exhibiting quantum coherence, most multiparticle subsystems exhibit classical properties of coherence. Consequently, the quantum dynamics within the macroscopic system are overshadowed by the classical coherence properties of the majority of the constituent multiphoton subsystems. The contributions from all these isolated subsystems lead to the classical properties of the open macroscopic system \cite{MaganaSciAdv}. Furthermore, the classical OAM correlations of the macroscopic system are reported in the first row of Figure \ref{Fig2}\textbf{b}. Sharing similarities with the classical macroscopic system, extracted multiparticle subsystems containing the same number of particles exhibit correlation peaks in their OAM distribution. As indicated by the second and third rows of Figure \ref{Fig2}\textbf{b}, the strength of these correlation peaks scales with the number of particles in the isolated subsystem. Here, the fourteen-particle subsystem $\left(n_{1}=7, n_{2}=7\right)$ shows stronger OAM correlations than its eight-particle counterpart $\left(n_{1}=4, n_{2}=4\right)$. Interestingly, the detection of multiphoton subsystems with different numbers of particles in each arm demonstrates the presence of OAM-correlation holes as indicated in the fourth and fifth rows of Figure \ref{Fig2}\textbf{b}. This first observation of correlation holes unveils the presence of quantum coherence governing interactions among certain many-body subsystems that constitute the classical macroscopic system \cite{HongNP2024, JeltesNature2007}. Remarkably, multiparticle subsystems characterized with classical properties of coherence ($\tilde{g}^{(2)}(\ell_1,\ell_2,n_1,n_2)>1$) produce many-body Hanbury Brown and Twiss interference when an angular double-slit structure is placed in the source \cite{MaganaSciAdv}. These measurements showing multiparticle anti-coalescence are reported in Figure \ref{Fig2}\textbf{c}. Moreover, the angular double-slit interference of the multiparticle subsystems exhibiting OAM-correlation holes ($\tilde{g}^{(2)}(\ell_1,\ell_2,n_1,n_2)<1$) produce the multiparticle coalescence reported in  Figure \ref{Fig2}\textbf{d} \cite{HiekkamakiPRL2021}. This effect is mediated by a multiparticle Hong-Ou-Mandel effect in subsystems with up to eleven particles. We now generalize this kind of interference process to demonstrate classical-to-quantum correspondences and anti-correspondences.

The effects discussed in Figure \ref{Fig2} lead to both ordinary and anomalous classical-to-quantum transitions \cite{ZurekPT, MartiniRMP, JaegerAJP, JarzynskiPRX2015, EstebanNatCom2012, SavageNature2012}. These transitions can be observed through the scattering of the macroscopic system by the angular double-slit structure shown in Figure \ref{Fig1}\textbf{c}. In this case, we project the scattered multiparticle subsystems in one of the arms into different OAM modes, from $\ell_1=-15$ to $\ell_1=15$, whereas the scattered subsystems in the second arm are projected into $\ell_2=3$. Given the presence of OAM correlations in classical light \cite{MaganaSciAdv}, the correlations between the two arms lead to the formation of an interference pattern with a peak around $\ell_1=3$, as shown in Figure \ref{Fig3}\textbf{a}. As stipulated by the correspondence principle, the constituent many-body subsystems produce interference patterns similar to those of the hosting system. Due to the anti-coalescence effects mediating these interference processes ($\tilde{g}^{(2)}(n_1,n_2)>1$), the patterns in Figure \ref{Fig3}\textbf{b} also exhibit a peak around $\ell_1=3$. These measurements indicate ordinary classical-to-quantum transitions \cite{ZurekPT, MartiniRMP, JaegerAJP}. Nevertheless, the extracted many-body systems with a larger number of particles produce quantum-enhanced visibilities which surpass that of the classical hosting system. Remarkably, the coalescence of multiparticle subsystems ($\tilde{g}^{(2)}(n_1,n_2)<1$) leads to the observation of anomalous classical-to-quantum correspondences. In deviation from the Planck-Bohr correspondence principle, the extracted many-body subsystems produce the anti-correlated interference structures reported in Figure \ref{Fig3}\textbf{c}. Surprisingly, these multiparticle interference patterns reveal a peak at $\ell_1=-3$. For nearly two decades, this behavior has been attributed to OAM-entangled photons \cite{JhaPRL}, which exhibit OAM anti-correlations due to the conservation of OAM in spontaneous parametric down-conversion processes \cite{mairnature2001}. Surprisingly, in our experiment, these nonclassical many-body subsystems showing classical-to-quantum anti-correspondences were extracted from an open macroscopic optical system. Notably, our theoretical model, which utilizes a complex-Gaussian probability density function, successfully captures all the features observed in our experiment.

While extensive research has been devoted to investigating the nonclassical properties of macroscopic quantum systems \cite{ZurekPT,MartiniRMP,ArndtNatPhys,VedralNature,GerlichNatComm,FeinNatPhys,DittelPRX,LvovskyNatPhys,BouganneNatPhys}, this work provides the first direct evidence of the quantum signatures of the classical macroscopic world. Our measurements of the quantum dynamics of an open macroscopic system demonstrate the contrasting nature between the so-called classical and quantum worlds \cite{ZurekPT,MartiniRMP,ArndtNatPhys,VedralNature,GerlichNatComm,FeinNatPhys,DittelPRX,LvovskyNatPhys,BouganneNatPhys,AspuruNatPhys,FrowisRMP,GeorgescuRMP,JaegerAJP}. Despite the prevalent acceptance of the correspondence principle in quantum physics, our observations reveal a lack of direct classical-to-quantum correspondence between classical macroscopic systems and their constituent many-body quantum subsystems. These effects were described using a novel quantum mechanical description of our macroscopic system through a complex-Gaussian probability function. As such, quantum physics allows for anomalous classical-to-quantum correspondences between a classical system and its constituent nonclassical subsystems. These results suggest that quantum physics permits exceptions to the Planck-Bohr formulation of the correspondence principle. Furthermore, our work outlines novel mechanisms to extract and isolate quantum systems from open classical systems \cite{WubsNatPhy2024, Fatemeh2024Imaging}. This possibility opens new paradigms in quantum physics with enormous implications for the development of robust quantum technologies \cite{WubsNatPhy2024}. We believe that our discoveries unveil universal properties of many-body systems that are relevant for diverse fields ranging from condensed matter to nuclear physics \cite{ZhangAOP2024, MartiniRMP, AspuruNatPhys, FrowisRMP, Nerenberg2024}.

\section*{Materials and Methods}

\section*{Source Preparation and experiment}

As shown in Figure \ref{Fig1}\textbf{c}, we utilized a continuous-wave laser operating at a wavelength of 532 nm. The initial beam is spatially cleaned and expanded using a 4-f system to match the size of the digital micromirror device (DMD, DLP LightCrafter 6500 Evaluation Module). In our experiment, we generate a partially coherent beam with engineered photon statistics. To experimentally generate partially coherent multiphoton states, we dynamically modulate the laser beam's spatial properties of coherence by displaying random phase screens generated through the Kolmogorov model of turbulence on the DMD \cite{HongLPR}. The Kolmogorov phase screen is given by $\Phi=\operatorname{Re}\left(\mathcal{F}^{-1}(\mathbb{M} \sqrt{\phi})\right)$, where $\mathbb{M}$ is the encoded random matrix, and the approximated power spectral density $\phi \approx 0.023 r_{0}^{-\frac{5}{3}} f^{-\frac{11}{3}}$. Here, $\operatorname{Re}(z)$ indicates the real part of a complex number $z$, $\mathcal{F}^{-1}$ is the inverse Fourier transform, and $f$ is the spatial frequency of the light field. Particularly, the encoding of Kolmogorov phase screens in a DMD is achieved through amplitude-only spatial light modulation. These generated Kolmogorov phase screens are encoded into a video at a rate of 60 frames per second and displayed on the DMD. Another 4-f system and an iris are used to filter the first diffraction order of the beam reflected off the DMD. We then prepare two partially coherent multiphoton sources by dividing the generated beam with a 50:50 beam splitter. These beams illuminate two angular double-slits on two spatial light modulators (SLMs). Each SLM also projects the scattered field into specific OAM values, and the beams reflected by the SLMs are weakly coupled into two single-mode fibers \cite{MaganaSciAdv}. These fibers direct photons to superconducting nanowire single-photon detectors (SNSPDs) that performs photon number resolving (PNR) detection \cite{omar2019Multiphoton, YouAPR2020, YouAPR2021}.

%\section{Acknowledgments}
%C. Y., F.M., J. F., and O. S. M. L. acknowledge support from the Army Research Office (ARO), through the Early Career Program (ECP) under the grant no. W911NF-22-1-0088. M. H. and O. S. M. L. thank the U.S. Department of Energy (DOE), Office of Science (SC), for supporting this research through the Program of Nuclear Physics under the NP-QIS grant: DE-SC0023694. Also, R. B. D. and O. S. M. L. acknowledge funding from the National Science Foundation through Grant No. OMA 2231387. R.J.L.-M. thankfully acknowledges financial support by DGAPA-UNAM under the project UNAM-PAPIIT IN101623. We thank Dr. Ivan Agullo and Dr. Javier Sanchez-Mondragon for discussions. 

\bibliography{main}% Produces the bibliography via BibTeX.

\clearpage

\onecolumngrid

\section*{Orbital angular momentum multiparticle correlations}

In this section, we describe the orbital angular momentum (OAM) multiparticle correlations in a classical random optical system. We aim to express our beam in the Laguerre–Gaussian (LG) mode set at $z=0$ given by
\begin{equation}
    \text{LG}^{\ell}_{p}(r,\phi) \equiv \text{LG}^{\ell}_{p}(r)\exp (i \ell \phi)= \sqrt{\frac{2 p !}{\pi(p+|\ell|) !}} \frac{1}{w_0}\left(\frac{r \sqrt{2}}{w_0}\right)^{|\ell|} \exp \left(-\frac{r^2}{w^2_0}\right) L_p^{|\ell|}\left(\frac{2 r^2}{w^2_0}\right) \exp (i \ell \phi),
\end{equation}
where $L_p^{|\ell|}(\cdot)$ are the associated Laguerre polynomials and $w_0$ is the beam width. In addition, $p$ is the radial index, $\ell$ is the azimuthal index, and $r$ and $\phi$ represent the spatial coordinate in the polar coordinate system. These LG modes form an orthonormal basis, so our random beam can naturally be decomposed into separate LG modes \cite{JhaPRL}. In particular, the type of random beam that we use in our experiment has the following coherent-mode representation
\begin{equation}\label{EqS2}
      W(r,\phi,r',\phi') =\braket{E^{(-)}(r,\phi)E^{(+)}(r',\phi')}=\mu^*(r,\phi)\mu(r',\phi')+ \sqrt{\bar{n}(r,\phi)\bar{n}(r',\phi')}g(r,\phi,r',\phi').
\end{equation}
Here, $\mu(r,\phi)$ denotes the average electric-field amplitude $\braket{E^{(+)}(r,\phi)}$ at point $(r,\phi)$, $\bar{n}(r,\phi) \propto \exp\left[-|r|^2/\sigma_0\right]$ represents the mean photon number at point $(r,\phi)$, and $g(r,\phi,r',\phi') \propto \exp\left[-|\mathbf{r}-\mathbf{r}'|^2/\sigma_1\right]$ is the normalized correlation term between the two points, where $\sigma_0$ and $\sigma_1$ are real, positive constants.
In our experiment, we project our random beam onto a particular value of $\ell$, which is equivalent to summing over all values of $p$ for that value of $\ell$. Therefore, for a given $\ell$, we will define the coherent-amplitude of that OAM mode by
\begin{equation}
    E^{(+)}_{\ell}= \sum_p\int r drd\phi \text{LG}^{*\ell}_{p}(r,\phi)E^{(+)}(r,\phi) \equiv \sum_{p} E^{(+)}_{\ell,p},
\end{equation}
such that the total coherent amplitude is given by $E^{(+)}(r,\phi) = \sum_{\ell,p} E^{(+)}_{\ell,p}\text{LG}^{\ell}_{p}(r,\phi)$. In this work, we are interested in studying the quantum correlations between photons in these OAM modes after the random beam has passed through an angular double-slit structure $S(\phi)=\text{rect}(\frac{\phi}{w})+\text{rect}(\frac{\phi-\phi_0}{w})$. Here, $\text{rect}(\cdot)$ is the Heaviside Pi function, $w$ represents the angular width of the slits, and $\phi_0$ defines the angular separation between the slits. Due to the inherent Gaussian-statistical nature of our random light source, we can obtain the distribution of OAM quanta directly from the classical probability distribution of coherent amplitudes $E^{(+)}_\ell$ \cite{dawkins2024quantum}. In order to obtain this distribution, we must first compute the mean  
\begin{equation}
    \mu_\ell = \braket{E^{(+)}_\ell} = \sum_p \int rdr d\phi\text{ } \text{LG}^{*\ell}_{p} (r,\phi)\mu(r,\phi) S(\phi),
\end{equation}
which is well-defined for $\mu(r,\phi)S(\phi) \equiv \braket{E^{(+)}(r,\phi)}$. In addition, we have 
\begin{equation}
\begin{aligned}
    \sigma_\ell &= \braket{E^{(-)}_\ell E^{(+)}_\ell} - \braket{E^{(-)}_\ell}\braket{E^{(+)}_\ell}\\ &= \sum_{p_1,p_2}\int rdrd\phi\int r'dr'd\phi' \text{LG}^{*\ell}_{p_1}(r,\phi)\text{LG}^{\ell}_{p_2}(r',\phi') g(r,\phi,r',\phi')\sqrt{\bar{n}(r,\phi)\bar{n}(r',\phi')}S(\phi)S(\phi'),
\end{aligned}
\end{equation}
where $g(r,\phi,r',\phi')\sqrt{\bar{n}(r,\phi)\bar{n}(r',\phi')}S(\phi)S(\phi')\equiv \braket{E^{(-)}(r,\phi)E^{(+)}(r',\phi')} - \braket{E^{(-)}(r,\phi)}\braket{E^{(+)}(r',\phi')}$ is the correlation function of the electric field after it has passed through the slit structure. It is useful to note here that this is a real quantity. Lastly, to study the correlations between two different OAM modes, we compute their cross-correlation function
\begin{equation}
\begin{aligned}
     \eta_{\ell_1,\ell_2} &= \braket{E^{(-)}_{\ell_1}E^{(+)}_{\ell_2}} - \braket{E^{(-)}_{\ell_1}}\braket{E^{(+)}_{\ell_2}} \\&= \sum_{p_1,p_2}\int r dr d\phi \int r' dr'd\phi' \text{ }\text{LG}^{*\ell_1}_{p_1}(r,\phi)\text{LG}^{\ell_2}_{p_2}(r',\phi') g(r,\phi,r',\phi')\sqrt{\bar{n}(r,\phi)\bar{n}(r',\phi')}S(\phi)S(\phi').
\end{aligned}
\end{equation}
This time, the resulting quantity will generally be complex-valued, having nonzero real and imaginary parts. Additionally, we note that $\sigma_\ell = \eta_{\ell,\ell}$. The total covariance matrix for a two-mode OAM system, written in terms of its real and imaginary components, is then given by
\begin{equation}
    \Gamma_{\ell_1,\ell_2} = \begin{pmatrix}
        \sigma_{\ell_1}&0&\text{Re}[\eta_{\ell_1,\ell_2}]&-\text{Im}[\eta_{\ell_1,\ell_2}]\\
        0&\sigma_{\ell_1}&\text{Im}[\eta_{\ell_1,\ell_2}]&\text{Re}[\eta_{\ell_1,\ell_2}]\\
        \text{Re}[\eta_{\ell_1,\ell_2}]&\text{Im}[\eta_{\ell_1,\ell_2}]&\sigma_{\ell_2}&0\\
        -\text{Im}[\eta_{\ell_1,\ell_2}]&\text{Re}[\eta_{\ell_1,\ell_2}]&0&\sigma_{\ell_2}
    \end{pmatrix}.
\end{equation}
From here, we can find the desired probability density via the formula
\begin{equation}
    P(\Vec{\boldsymbol{r}}) = \frac{1}{4\pi^2 \sqrt{|\boldsymbol{\Gamma}|}}e^{-\frac{1}{2}\left(\Vec{\boldsymbol{r}}-\Vec{\boldsymbol{\mu}}\right)^T\boldsymbol{\Gamma}^{-1}\left(\Vec{\boldsymbol{r}}-\Vec{\boldsymbol{\mu}}\right)}
\end{equation}
for a real-Gaussian random 4-vector $\Vec{\boldsymbol{r}}$. Writing $E^{(+)}_{\ell_1} = \alpha$ and $E^{(+)}_{\ell_2} = \beta$ for simplicity, we obtain
\begin{equation}
    P_{\ell_1,\ell_2}(\alpha,\beta) = \frac{1}{4\pi^2 \big|\sigma_{\ell_1}\sigma_{\ell_2} - |\eta_{\ell_1,\ell_2}|^2\big|} \exp\left[-\frac{\sigma_{\ell_2}|\alpha - \mu_{\ell_1}|^2 + \sigma_{\ell_1}|\beta - \mu_{\ell_2}|^2 - 2\text{Re}\left[(\alpha^*-\mu_{\ell_1}^*)(\beta-\mu_{\ell_2})\eta_{\ell_1,\ell_2}\right]}{2(\sigma_{\ell_1}\sigma_{\ell_2} - |\eta_{\ell_1,\ell_2}|^2)}\right].
\end{equation}
Then, utilizing the natural isomorphism between a classical complex electric field and a quantum coherent state, we conclude that the density matrix describing the quantum-mechanical behavior of this system is given by
\begin{equation}
    \hat{\rho}_{\ell_1,\ell_2} = \int d^2\alpha d^2\beta P_{\ell_1,\ell_2}(\alpha,\beta)|\alpha,\beta\rangle\langle\alpha,\beta|.
\end{equation}
%It is important to note here that this formula is specifically for the case where $\ell_1 \neq \ell_2$. In the case that the two OAM values are equal, the probability density function would reduce to a function of only one complex amplitude. The resulting quantum state would then have just a single mode. In either case, to experimentally access the correlation dynamics between these two modes, we must send the state through a beam-splitter. 
Notably, the function $P_{\ell_1,\ell_2}(\alpha,\beta)$ represents the correlations between two different modes of OAM. When $\ell_1 = \ell_2$, the state reduces to that of a single mode. In order to experimentally access the multiparticle correlations between two different OAM modes, we will first send this state through a beam-splitter (with beam-splitter angle $\theta$), and afterwards we will project each arm onto OAM modes $\ell_1$ and $\ell_2$. The resulting state after these steps will be
\begin{equation}
    \hat{\rho}_{\ell_1,\ell_2} = \int d^2\alpha d^2\beta P_{\ell_1,\ell_2}(\alpha,\beta)|\alpha\cos(\theta),i\beta\sin(\theta)\rangle\langle\alpha\cos(\theta),i\beta\sin(\theta)|.
\end{equation}

\begin{figure*}[!ht]
  \centering
 \includegraphics[width=0.75\textwidth]{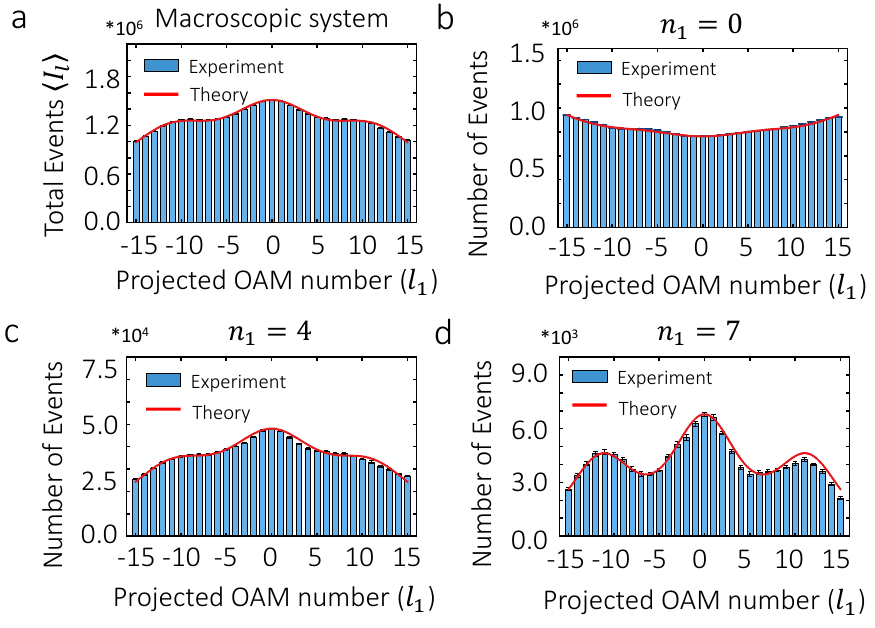}
\caption{\textbf{Macroscopic and multiphoton quantum scattering of a single arm.} The interference pattern in \textbf{a} shows the OAM-dependence for a single mode of our macroscopic system with a double-slit. Here, the visibility of these fringes is determined by the coherence of our source. In panel \textbf{b}, we show the OAM-dependence for zero-photon detections. In contrast to the macroscopic system, this interference pattern exhibits a dip centered at $\ell_1=0$. Panels \textbf{c} and \textbf{d} show the OAM-dependent interference fringes for the detection of 4-photon and 7-photon wavepackets, respectively. Similar to panel \textbf{a}, these interferences resemble that of the macroscopic system. Notably, multiphoton subsystems with a larger number of photons have high-visibility fringes.}
\label{FigS2}
\end{figure*}

By taking the expectation value of the Fock projection operators $\ket{N,M}\bra{N,M}$ with respect to this state, we can predict the multiphoton correlation statistics between different OAM modes. However, in order to calculate this, we will need to compute $\mu_\ell$ and $\eta_{\ell_1,\ell_2}$, and the corresponding $\text{Tr}\big[\hat{\rho}_{\ell_1,\ell_2}\ket{N,M}\bra{N,M}\big]$. Starting with $\mu_\ell$, we note that the function $\mu(r,\phi)$ has a Gaussian profile. In other words, $\mu(r,\phi)\equiv \text{LG}_0^0(r,\phi)$. Therefore, we can write
\begin{equation}
\begin{aligned}
    \mu_\ell &= \bigg[\sum_p \int r dr \text{LG}^{*\ell}_p(r)\text{LG}^0_0(r)\bigg]\int d\phi e^{-i \phi \ell} S(\phi)\\
    &= 2w e^{-i\phi_0 \ell/2}\cos\bigg(\frac{\phi_0 \ell}{2}\bigg)\sinc\bigg(\frac{w \ell}{2}\bigg),
\end{aligned}
\end{equation}
where the last line follows from the identity $\int rdr \text{LG}^{*\ell}_p(r)\text{LG}^0_0(r) = \delta_{p0}$. Next, in order to compute $\eta_{\ell_1,\ell_2}$, we will first expand $g(r,\phi,r',\phi')\sqrt{\bar{n}(r,\phi)\bar{n}(r',\phi')}$ into the OAM-mode basis as
\begin{equation}
    g(r,\phi,r',\phi')\sqrt{\bar{n}(r,\phi)\bar{n}(r',\phi')} = \sum_{\ell,p,p'} \alpha_{\ell,p,p'} \text{LG}^{\ell}_p(r,\phi)\text{LG}^{*\ell}_{p'}(r',\phi'),
\end{equation}
for some coefficients $\alpha_{\ell,p,p'}$. With this, we can write
\begin{equation}
\begin{aligned}
    \eta_{\ell_1,\ell_2} &= \sum_{p_1,p_2}\sum_{\ell,p,p'} \alpha_{\ell,p,p'}\int r dr d\phi \int r' dr'd\phi' \text{ }\text{LG}^{*\ell_1}_{p_1}(r,\phi)\text{LG}^{\ell_2}_{p_2}(r',\phi')\text{LG}^{\ell}_p(r,\phi)\text{LG}^{*\ell}_{p'}(r',\phi')S(\phi)S(\phi')\\
    &= \sum_{p_1,p_2}\sum_{\ell,p,p'} \alpha_{\ell,p,p'} \frac{\delta_{pp_1}\delta_{p'p_2}}{4\pi^2} \int d\phi d\phi' S(\phi)S(\phi') \exp[-i\phi(\ell_1-\ell) - i\phi'(\ell - \ell_2)]\\
    &\equiv \frac{w^2}{\pi^2}e^{-i\phi_0(\ell_1-\ell_2)/2}\sum_\ell C_\ell \cos\left(\frac{\phi_0(\ell_1-\ell)}{2}\right)\cos\left(\frac{\phi_0(l-\ell_2)}{2}\right)\sinc\left(\frac{w(\ell_1-\ell)}{2}\right)\sinc\left(\frac{w(\ell-\ell_2)}{2}\right),
\end{aligned}
\end{equation}
where here we have defined $C_\ell = \sum_{p,p'}\alpha_{\ell,p,p'}$. The coefficients $C_\ell$ represent the contribution of each OAM mode. While an explicit computation of $C_\ell$ would be fairly complicated to carry out, it is commonly approximated by $C_\ell\approx e^{-\ell^2/\lambda}$ for some OAM-bandwidth $\lambda$. Therefore, the $\eta_{\ell_1,\ell_2}$ terms can be evaluated through a numerical summation over $\ell$. We then have
\begin{equation}
    \eta_{l_1,l_2}\approx \eta_0 e^{-i\phi_0(\ell_1-\ell_2)/2}\cos\left(\frac{\phi_0(\ell_1-\ell_2)}{2}\right)\sinc\left(\frac{w(\ell_1-\ell_2)}{2}\right)\text{exp}\left(-\frac{(\ell_1+\ell_2)^2}{4\lambda}\right)
\end{equation}
form some $\eta_0$ when $\lambda$ is large. Now, having obtained analytical expressions for each term in the density matrix's P-function, all that's left is to compute its individual matrix elements. This can be accomplished using the methods described in Ref. \cite{dawkins2024quantum}. With this, we have successfully broken the the state up into its constituent multiphoton subsystems, and are thus able to study their quantum correlation statistics. This is accomplished through use of the multiphoton correlation function
\begin{equation}
    \tilde{g}^{(2)}(\ell_1,\ell_2,N,M) = \frac{\text{Tr}\big[\hat{\rho}_{\ell_1,\ell_2}\ket{N,M}\bra{N,M}\big]}{\left(\sum_n \text{Tr}\big[\hat{\rho}_{\ell_1,\ell_2}\ket{n,M}\bra{n,M}\big]\right)\left(\sum_m \text{Tr}\big[\hat{\rho}_{\ell_1,\ell_2}\ket{N,m}\bra{N,m}\big]\right)}.
\end{equation}
As expected, when $\ell_1 = \ell_2$, the subsystems are strongly correlated for $N = M$ and strongly anticorrelated for $N \ll M$ or $N \gg M$. When $\ell_1$ and $\ell_2$ are very different from one-another, however, the correlation function becomes approximately $1$.

Our model, as described above, also enables us to describe the classical first-order interference pattern produced by our classical random optical system. The measured pattern together with the theoretical prediction from this model is shown in Figure \ref{FigS2}\textbf{a}. As expected, this pattern is centered at $\ell_{1}=0$ with a visibility dictated by the coherence of the macroscopic system. Notably, the projection of this macroscopic system into its quantum constituents unveils different interaction processes. For example, in Figure \ref{FigS2}\textbf{b}, we show that the measurement of vacuum wavepackets leads to a dip at $\ell_{1}=0$. In Figures \ref{FigS2}\textbf{c-d}, we show similar patterns produced by additional constituent wavepackets containing 4 and 7 photons, respectively. These patterns resemble that of the macroscopic system, with the difference being that wavepackets with a higher number of photons exhibit a higher visibility. 

\section*{Explicit derivation of the correlation functions}

In this section, we provide explicit derivations for the equations which describe the correlations presented in the main body of our paper. We note that to obtain the corresponding expressions for the single-slit case, we simply remove the second term in $S(\phi)$. The first formula which we will derive is that of the classical, macroscopic correlation function, given by
\begin{equation}
    g^{(2)}(\ell_1,\ell_2) = \frac{\text{Tr}\left[\hat{\rho}_{\ell_1,\ell_2}\hat{n}_1\hat{n}_2\right]}{\text{Tr}\left[\hat{\rho}_{\ell_1,\ell_2}\hat{n}_1\right]\text{Tr}\left[\hat{\rho}_{\ell_1,\ell_2}\hat{n}_2\right]}.
\end{equation}
It is fairly straightforward to compute each of the terms in this expression. Specifically, the numerator is given by
\begin{equation}
    \begin{aligned}
        \text{Tr}\left[\hat{\rho}_{\ell_1,\ell_2}\hat{n}_1\hat{n}_2\right] &= \int d^2\alpha d^2\beta P_{\ell_1,\ell_2}(\alpha,\beta)|\alpha|^2\cos^2(\theta)|\beta|^2\sin^2(\theta)\\
        &= \bigg[(2\sigma_{\ell_1}+|\mu_{\ell_1}|^2)(2\sigma_{\ell_2}+|\mu_{\ell_2}|^2) + 4|\eta_{\ell_1\ell_2}|^2 + 4 \text{Re}\left[\mu^*_{\ell_1}\mu_{\ell_2}\eta_{\ell_1\ell_2}\right]\bigg]\cos^2(\theta)\sin^2(\theta),
    \end{aligned}
\end{equation}
while the two denominator terms are given by
\begin{equation}
    \begin{aligned}
        \text{Tr}\left[\hat{\rho}_{\ell_1,\ell_2}\hat{n}_1\right] &= \int d^2\alpha d^2\beta P_{\ell_1,\ell_2}(\alpha,\beta)|\alpha|^2\cos^2(\theta)\\
        &= \bigg[2\sigma_{\ell_1}+|\mu_{\ell_1}|^2\bigg]\cos^2(\theta),\\
        \text{Tr}\left[\hat{\rho}_{\ell_1,\ell_2}\hat{n}_2\right] &= \int d^2\alpha d^2\beta P_{\ell_1,\ell_2}(\alpha,\beta)|\beta|^2\sin^2(\theta)\\
        &= \bigg[2\sigma_{\ell_2}+|\mu_{\ell_2}|^2\bigg]\sin^2(\theta).
    \end{aligned}
\end{equation}
As a result, we obtain the correlation function
\begin{equation}
    g^{(2)}(\ell_1,\ell_2) = 1 + \frac{4|\eta_{\ell_1\ell_2}|^2 + 4 \text{Re}\left[\mu^*_{\ell_1}\mu_{\ell_2}\eta_{\ell_1\ell_2}\right]}{(2\sigma_{\ell_1}+|\mu_{\ell_1}|^2)(2\sigma_{\ell_2}+|\mu_{\ell_2}|^2)}.
\end{equation}
In its current form, the resulting interference pattern is not especially clear. However, we recall from the last section that
\begin{equation}
    \begin{aligned}
        \mu_\ell &= \mu_0 e^{-i\phi_0 \ell/2}\cos\bigg(\frac{\phi_0 \ell}{2}\bigg)\sinc\bigg(\frac{w \ell}{2}\bigg),\\
        \eta_{\ell_1,\ell_2} &\approx \eta_0 e^{-i\phi_0(\ell_1-\ell_2)/2}\cos\left(\frac{\phi_0(\ell_1-\ell_2)}{2}\right)\sinc\left(\frac{w(\ell_1-\ell_2)}{2}\right)\text{exp}\left(-\frac{(\ell_1+\ell_2)^2}{4\lambda}\right),
    \end{aligned}
\end{equation}
for some constants $\mu_0,\eta_0$, and where the spiral bandwidth $\lambda$ was assumed to be large. Furthermore, we have that $\sigma_\ell = \eta_{\ell\ell}$. These formulae imply that, for particular OAM projections, the statistics of the resulting state may become perfectly thermal. While the spatial correlation statistics of our source can be estimated very well by Eq. (\ref{EqS2}), this is still only an approximation. Specifically, we find that the coherent amplitude $|\mu_\ell|^2$ does not have the predicted perfect visibility. This could be a result of our beam's not being perfectly decomposable into the sum of a thermal Gaussian beam and a coherent Gaussian beam. This issue can be corrected for by adding a constant to the coherent amplitude which represents the deviation of our source's statistical decomposition from the model. The corrected $\mu_l$ can then be written as
\begin{equation}
    \mu_\ell \equiv \mu_0\bigg[(1-\zeta)e^{-i\phi_0 \ell/2}\cos\bigg(\frac{\phi_0 \ell}{2}\bigg)\sinc\bigg(\frac{w \ell}{2}\bigg) + \zeta\bigg],
\end{equation}
where here $0\leq\zeta\leq 1$ represents a correction term. This, together with our approximation for $\eta_{\ell_1,\ell_2}$, allows for the computation of $g^{(2)}(\ell_1,\ell_2)$.

Next, to perform post-selective measurements on our source and, consequently, to compute second-order multiphoton coherence, we must calculate
\begin{equation}\label{eqS9}
    \begin{aligned}
\text{Tr}\left[\hat{\rho}_{\ell_1,\ell_2}\ket{N,M}\bra{K,L}\right] &= \int d^2\alpha d^2\beta P_{\ell_1,\ell_2}(\alpha_1,\alpha_2,\beta_1,\beta_2)\text{exp}\left[-|\alpha|^2-|\beta|^2\right]\frac{\alpha^N\alpha^{*K}\beta^M\beta^{*L}}{\sqrt{N!K!M!L!}}i^{M-L}\cos^{N+K}(\theta)\sin^{M+L}(\theta)\\
        &=\frac{1}{\sqrt{N!K!M!L!}}\int d^2\alpha d^2\beta P_{\ell_1,\ell_2}(\alpha_1,\alpha_2,\beta_1,\beta_2)\text{exp}\left[-\alpha_1^2-\alpha_2^2-\beta_1^2-\beta_2^2\right]\\
        &\text{ }\text{ }\text{ }\text{ }\text{ }\text{ }\text{ }\text{ }\text{ }\text{ }\text{ }\text{ }\text{ }\text{ }\text{ }\text{ }\text{ }\text{ }\text{ }\text{ }\text{ }\text{ }\text{ }\text{ }\text{ }\text{ }\text{ }\text{ }\text{ }\text{ }\text{ }\text{ }\text{ }\text{ }\text{ }\times(\alpha_1+i\alpha_2)^N(\alpha_1-i\alpha_2)^K(\beta_1+i\beta_2)^M(\beta_1-i\beta_2)^L\\
        &= \frac{1}{\sqrt{N!K!M!L!}}\sum_{n=0}^N\sum_{k=0}^K\sum_{m=0}^M\sum_{l=0}^L\binom{N}{n}\binom{K}{k}\binom{M}{m}\binom{L}{l} i^{n-N+m-M-k+K-l+L} \\
        &\text{ }\text{ }\text{ }\text{ }\times \int d\alpha_1d\alpha_2d\beta_1d\beta_2 P_{\ell_1,\ell_2}(\alpha_1,\alpha_2,\beta_1,\beta_2)e^{-\alpha_1^2\cos^2(\theta)-\alpha_2^2\cos^2(\theta)-\beta_1^2\sin^2(\theta)-\beta_2^2\sin^2(\theta)}\\
        &\text{ }\text{ }\text{ }\text{ }\times\alpha_1^{n+k}\alpha_2^{N+K-n-k}\beta_1^{m+l}\beta_2^{M+L-m-l} i^{M-L}\cos^{N+K}(\theta)\sin^{M+L}(\theta),
        \end{aligned}
\end{equation}
where here we have written $\alpha \equiv \alpha_1 + i \alpha_2$ and $\beta \equiv \beta_1 + \beta_2$. Then, defining $\mu_{l_1}\equiv \mu_1 + i\mu_2$, $\mu_{l_2} \equiv \mu_3 + i\mu_4$, and $\eta_{l_1,l_2}\equiv \eta_1+i\eta_2$, as well as $\sigma_{l_i}\equiv\sigma_i$ for simplicity. To evaluate the integral, we will perform the following change of coordinates:
\begin{equation}
    \begin{aligned}
        \alpha_1 &= A_1 + \frac{\mu_1 x_2}{\cos^2(\theta)+x_2},\\
        \alpha_2 &= A_2 + \frac{\mu_2 x_2}{\cos^2(\theta)+x_2},\\
        \beta_1 &= \frac{y_1}{2(x_1+1)}A_1 + \frac{y_2}{2(x_1+1)}A_2 - \frac{y_2}{2(x_1+1)}B_1 + \frac{\mu_3 x_1}{x_1+1},\\
        \beta_2 &= -\frac{y_2}{2(x_1+1)}A_1 + \frac{y_1}{2(x_1+1)}A_2 - \frac{y_1}{2(x_1+1)}B_2 + \frac{\mu_4 x_1}{x_1+1},
    \end{aligned}
\end{equation}
where here we have defined the intermediate variables
\begin{equation}
    \begin{aligned}
        x_i &= \frac{\sigma_i}{2(\sigma_1\sigma_2-\eta_1^2-\eta_2^2)},\\
        y_i &= \frac{\eta_i}{2(\sigma_1\sigma_2-\eta_1^2-\eta_2^2)}.
    \end{aligned}
\end{equation}
Then, by defining the further-abstracted variables
\begin{equation}
    \begin{aligned}
        z_0 &= (1+x_2) - \frac{y_1^2 + y_2^2}{4(1+x_1)},\\
        z_1 &= y_2^2,\\
        z_2 &= y_1^2,\\
        u_1 &= \frac{y_1\mu_3 - y_2\mu_4}{x_1+1} + \frac{\mu_1(y_1^2+y_2^2)}{2(1+x_1)(1+x_2)},\\
        u_2 &= \frac{y_1\mu_4 + y_2\mu_3}{x_1+1} + \frac{\mu_2(y_1^2+y_2^2)}{2(1+x_1)(1+x_2)},\\
        v_1 &= \frac{y_1\mu_1 + y_2\mu_2}{1+x_2},\\
        v_2 &= \frac{y_1\mu_2-y_2\mu_1}{1+x_2},\\
        C &= (\mu_1^2+\mu_2^2)\frac{x_2}{1+x_2} + (\mu_3^2 + \mu_4^2)\frac{x_1}{1+x_1}-\frac{y_1(\mu_1\mu_3+\mu_2\mu_4)+y_2(\mu_2\mu_3+\mu_1\mu_4)}{(1+x_1)(1+x_2)},
    \end{aligned}
\end{equation}
we can greatly simplify the exponential term of our integrand. Specifically, the Gaussian correlation terms are decoupled in this new coordinate frame. Our goal is now to evaluate the integral of general form
\begin{equation}
    f_{\ell_1,\ell_2}(n,m,k,l) = \int d\alpha_1d\alpha_2d\beta_1d\beta_2 P_{\ell_1,\ell_2}(\alpha_1,\alpha_2,\beta_1,\beta_2)e^{-\alpha_1^2-\alpha_2^2-\beta_1^2-\beta_2^2}\alpha_1^{n}\alpha_2^{m}\beta_1^{k}\beta_2^{l}.
\end{equation}
In the decoupled coordinate frame, this integral becomes
\begin{equation}
    \begin{aligned}
        f_{\ell_1,\ell_2}(n,m,k,l) =& \sum_{a = 0}^n \sum_{b=0}^m \sum_{c_3 = 0}^{k-c_1-c_2}\sum_{c_2 = 0}^{k-c_1}\sum_{c_1 = 0}^k\sum_{d_3 = 0}^{l-d_1-d_2}\sum_{d_2 = 0}^{l-d_1}\sum_{d_1 = 0}^l \binom{n}{a}\binom{m}{b}\binom{k}{c_1\text{ }c_2\text{ }c_3}\binom{l}{d_1\text{ }d_2\text{ }d_3}\\
        &\times \left[\frac{\mu_1 x_2}{1+x_2}\right]^{n-a}\left[\frac{\mu_2 x_2}{1+x_2}\right]^{m-b}\left[\frac{y_1}{2(1+x_1)}\right]^{c_1}\left[\frac{y_2}{2(1+x_1)}\right]^{c_2}\left[-\frac{y_2}{2(1+x_1)}\right]^{c_3}\\&\times\left[\frac{\mu_3 x_1}{1+x_1}\right]^{k-c_1-c_2-c_3}\left[-\frac{y_2}{2(1+x_1)}\right]^{d_1}\left[\frac{y_1}{2(1+x_1)}\right]^{d_2}\left[-\frac{y_1}{2(1+x_1)}\right]^{d_3}\left[\frac{\mu_4 x_1}{1+x_1}\right]^{l-d_1-d_2-d_3}\\
        &\times \int dA_1dA_2dB_1dB_2 A_1^{a+c_1+d_1}A_2^{b+c_2+d_2}B_1^{c_3}B_2^{d_3}\\&\times\text{exp}\left[-z_0A_1^2 - z_0A_2^2 - z_1B_1^2-z_2B_2^2 - u_1A_1-u_2A_2-v_1B_1-v_2B_2-C\right],
    \end{aligned}
\end{equation}
where here we have used the multinomial theorem, with
\begin{equation}
    \binom{a}{b\text{ }c\text{ }d} = \frac{a!}{b!c!d!(a-b-c-d)!}.
\end{equation}
Then, using the Gaussian integral formula
\begin{equation}
\begin{aligned}
    I(n,a,b) &= \int dx \text{ }x^n \text{exp}\left[-a x^2 - b x\right]\\
    &= \frac{1}{2a^{1 + n/2}}\bigg[(-1 + (-1)^n)b\Gamma\left[1+\frac{n}{2}\right]\text{}_1F_1\left(1+\frac{n}{2};\frac{3}{2};\frac{b^2}{4a}\right)+(1+(-1)^n)\sqrt{a}\Gamma\left[\frac{1+n}{2}\right]\text{}_1F_1\left(\frac{1+n}{2};\frac{1}{2};\frac{b^2}{4a}\right)\bigg],
\end{aligned}
\end{equation}
we obtain
\begin{equation}
    \begin{aligned}
        f_{\ell_1,\ell_2}(n,m,k,l) =& \sum_{a = 0}^n \sum_{b=0}^m \sum_{c_3 = 0}^{k}\sum_{c_2 = 0}^{k-c_3}\sum_{c_1 = 0}^{k-c_2-c_3}\sum_{d_3 = 0}^{l}\sum_{d_2 = 0}^{l-d_2}\sum_{d_1 = 0}^{l-d_2-d_3} \binom{n}{a}\binom{m}{b}\binom{k}{c_1\text{ }c_2\text{ }c_3}\binom{l}{d_1\text{ }d_2\text{ }d_3}\\
        &\times \left[\frac{\mu_1 x_2}{1+x_2}\right]^{n-a}\left[\frac{\mu_2 x_2}{1+x_2}\right]^{m-b}\left[\frac{y_1}{2(1+x_1)}\right]^{c_1}\left[\frac{y_2}{2(1+x_1)}\right]^{c_2}\left[-\frac{y_2}{2(1+x_1)}\right]^{c_3}\\&\times\left[\frac{\mu_3 x_1}{1+x_1}\right]^{k-c_1-c_2-c_3}\left[-\frac{y_2}{2(1+x_1)}\right]^{d_1}\left[\frac{y_1}{2(1+x_1)}\right]^{d_2}\left[-\frac{y_1}{2(1+x_1)}\right]^{d_3}\left[\frac{\mu_4 x_1}{1+x_1}\right]^{l-d_1-d_2-d_3}\\
        &\times I(a+c_1+d_1,z_0,u_1)I(b+c_2+d_2,z_0,u_2)I(c_3,z_1,v_1)I(d_3,z_2,v_2)e^{-C},
    \end{aligned}
\end{equation}
and finally
\begin{equation}
    \begin{aligned}
        \text{Tr}\left[\hat{\rho}_{\ell_1,\ell_2}\ket{N,M}\bra{K,L}\right] &=\frac{1}{\sqrt{N!K!M!L!}}\sum_{n=0}^N\sum_{k=0}^K\sum_{m=0}^M\sum_{l=0}^L\binom{N}{n}\binom{K}{k}\binom{M}{m}\binom{L}{l} i^{n-N+m-M-k+K-l+L} \\
        &\text{ }\text{ }\text{ }\text{ }\times f_{\ell_1,\ell_2}(n+k,N+K-n-k,m+l,M+L-m-l)\\
        &\text{ }\text{ }\text{ }\text{ }\times i^{M-L}\cos^{N+K}(\theta)\sin^{M+L}(\theta).
    \end{aligned}
\end{equation}
With this formula, we can compute the joint photon number distribution of our source.

\end{document}